\documentclass[english,aps,pre,manuscript,twocolumn,10pt]{revtex4}
\usepackage[utf8]{inputenc}
\usepackage{amsmath}
\usepackage{amsthm}
\usepackage{amssymb}
\usepackage{cancel}
\usepackage{graphicx}
\makeatother

\begin{document}


\title
{ The correspondence between the Adam-Gibbs and the Rosenfield relations}

\author{Bidhan~Chandra~Bag}
\email{bidhanchandra.bag@visva-bharati.ac.in}
\affiliation{Department of Chemistry,\
	Visva-Bharati, Santiniketan,\
	India, 731235 }

\date{\today}

\begin{abstract}
In this paper, we derive both the Adam-Gibbs and the Rosenfield relations from the microscopic point of view and compare them with the numerical calculation for one and two dimensional systems. The comparison shows there is an excellent agreement between theoretical and numerical calculations for their valid zones (in terms of the thermodynamic temperature) as suggested by experiments. It implies that there may be a transition temperature at which the two relations correspond to each other. We derive a relation to calculate it. Then, we generalize the Rosenfield relation for configurational thermodynamic entropy-like quantity(TELQ) and time-dependent Shannon information entropy. At the same time, using a description with a fictitious Hamiltonian, we show that time-dependent configurational Shannon information entropy for a thermodynamic system (of Brownian particles) which is characterized by the absolute temperature, can not be recognized as thermodynamic entropy. At best, it can be identified as a thermodynamic entropy-like quantity. Furthermore, the description based on the fictitious Hamiltonian may lead to the conclusion that the correspondence between the Shannon information entropy and thermodynamic entropy is not a singular feature at equilibrium. It may be a continuation of the correspondence between
the information entropy and the thermodynamic entropy-like quantity. Thus, the present study appears to offer important justification for the postulate that the Shannon entropy at steady state may be regarded as a thermodynamic entropy. This postulate holds significant importance in the framework of stochastic thermodynamics.  
\end{abstract}

\maketitle

\section{Introduction}
Particles in a fluid move in space through a diffusion mechanism. At the equilibrium state of the fluid, we expect that there is a relation between the diffusion coefficient and the entropy of the system. In this context, the Adam-Gibbs and the Rosenfield relations are well-known in the literature. In Refs. \cite{bagchi1,bagchi2,bagchi3,liao}, one may find very good summaries about these relations. Considering a potential energy field with randomly distributed ruggedness, a microscopic theory was developed in \cite {bagchi2} to account for the Rosenfield
relation. An alternative approach in this regard was presented in Ref. \cite {liao}, calculating a thermodynamic entropy-like quantity, which depends on the viscosity of the medium in which non-interacting Brownian particles (infinite in number) comprise a thermodynamic system.

The Adam-Gibbs relation seems to work at low temperatures. The other relation is good for intermediate to high temperatures. One may expect the there may bay be a correspondence between the two. To the best of our
knowledge, this issue seems to be an open problem. It may be due to lacking of the microscopic derivation of the Adam-Gibbs relation. 
The major objective of the present study is
microscopic derivation of the Rosenfield and the Adam-Gibbs relations and to find the condition at which they may correspond to each other. In this context, we consider the Brownian motion
of a particle in a periodic potential. For this sytem, we calculate configurational thermodynamic entropy and derive the relations. Using these relations, we find a relation to calculate the temperature at which the transition between the relations may occur. Then, we address the Rosenfield-like relations for the configurational information entropy and thermodynamic entropy-like quantity. In all these relations, a constant b appears. It is given by, $b=\frac{2}{d}$ for a $d$ dimensional system.  

\section{Microscopic derivation of the relation between the thermodynamic entropy and the diffusion coefficient}
  Fluid molecules may  interact with each other. As a signature of that, their diffusion may be different compared  to the ideal gas situation. To understand the diffusion constant in the nonideal situation from microscopic point of view, either periodic or random potential energy fileds were considered in the literature \cite{bagchi1,bagchi2,bagchi3,lj,zwang,frau,sastry,kl,kl1,kl2,lat,jaku,ben,meng}. The potential energy field seems to be due to a qualitatively mean field approximation to account for the interaction among the particles. For the present purpose,      
we also consider that a Brownian particle is moving in a periodic potential energy field. The relevant Langevin equation of motion can be written as
 
\begin{equation}
\frac{dq}{dt}= -V^\prime/\eta +\xi(t)/\eta 
\label{eqm1}
\end{equation}
\noindent
Here, $V(q)$ is an effective periodic potential with period L,

\begin{equation}
V(q+L)=V(q)     \; \; \;, 
\label{eqm2}
\end{equation}
\noindent
and
\begin{equation}
V(q)=V_0 \cos (2\pi q/L)    \; \; \;.   
\label{eqm3}
\end{equation}
\noindent
$\eta$ in Eq.(\ref{eqm1}) is the viscosity coefficient, and it is related to the Gaussian fluctuating force, $\xi(t)$ by the fluctuation-dissipation as

\begin{equation}
\langle\xi(t)\xi(t^\prime)\rangle = 2k_BT\eta \delta(t-t^\prime) 
\label{eqm4}
\end{equation}

\noindent
Here, $k_B$ is the Boltzmann constant, and $T$ is the temperature of the thermal bath. However, the Fokker-Plack equation corresponding to Eq.(\ref{eqm1}) is given by

\begin{equation}
\frac{\partial \hat{f}(q,t)}{\partial t}=\frac{1}{\eta}\frac{\partial V^\prime(q) \hat{f}(q,t)}{\partial q}+\frac{T}{\eta} \frac{\partial^2 \hat{f}(q,t)}{\partial q^2}
\label{eqm5}
\end{equation}
\noindent
Here, we have used $k_B=1$. $\hat{f}(q,t)$ is the probability density and it approaches zero for $q=\pm\infty$. Now, one may define the reduced probability $f(q,t)$ from this as

\begin{equation}
f(q,t)=\sum_{n=-\infty}^{n=\infty}  \hat{f}(q+nL,t) 
\label{eqm5a}
\end{equation}
\noindent
Since the Fokker-Planck equation (\ref{eqm5}) is a linear one, then $f(q,t)$ also satisfies it, i.e.,

\begin{equation}
\frac{\partial f(q,t)}{\partial t}=\frac{1}{\eta}\frac{\partial V^\prime(q) f(q,t)}{\partial q}+\frac{T}{\eta} \frac{\partial^2 f(q,t)}{\partial q^2}
\label{eqm5b}
\end{equation}
\noindent
It is to be noted here that if the original problem(\ref{eqm1}) is extended over the entire real axis (with the natural boundary conditions, i.e., $\hat{f}(q,t)$ approaches zero for $q=\pm\infty$), then the probability density $\hat{f}(q,t)$ will never approach a meaningful stationary state. In other words,  $\hat{f}(q,t) \rightarrow 0$ at long time does not admit any further conclusions and is therefore not considered as a meaningful equilibrium state. But it is only the reduced density $f(q,t)$, associated with periodic boundary conditions, that tends towards a meaningful $t$-independent long-time limit. In particular, only after this mapping, we get the expected zero current at long times. 

Now, to avoid any confusion, we mention that to study the dynamics of Brownian particles in a periodic potential, the approximate time-dependent solution of Eq.(\ref{eqm5}) at long time was determined as \cite{eli} 

\begin{equation}
\hat{f}(q,t) \approx \frac{e^{-\frac{V(q)}{T}}e^{-\frac{q^2}{4D_{eff}t}}}{N_t} 
\label{eqm5c}
\end{equation}
\noindent
with $N_t=\sqrt{\frac{4\pi T t}{\eta}}$. $D_{eff}$ is the effective diffusion constant and it is given by Lifson and Jackson \cite{lj}, as

\begin{equation}
D_{eff}=\frac{D}{\langle e^{-V(q)/T}\rangle_L\langle e^{-V(q)/T}\rangle_L}  \; \; \;.
\label{eqm5d}
\end{equation}

\noindent
and
\begin{equation}
 {\langle e^{-V(q)/T}\rangle}_L=\frac{1}{L}\int_{-L/2}^{L/2} e^{-V(q)/T}dq  \; \; \;.
 \label{eqm7}
\end{equation}

\noindent
Here, $D=\frac{T}{\eta}$ is the diffusion constant for free particle. As a signature of non linear unbounded system, the above distribution
function is intuitively expected at long time. At high temperature limit, the Boltzmann factor tends to unity and $D_{eff}$ becomes $D$ and
thus $\hat{f}(q,t)$ reduces to the solution of the Einsteine's diffusion equation. Still one may confuse mathmatically with $\hat{f}(q,t)$(\ref{eqm5c}) due to $D_{eff}$ as well as the nature of the averaging in Eq.(\ref{eqm5d}) which does not involve $\hat{f}(q,t)$. The distribution funtion was determined by matching between the two distribution functions,  $\hat{f}(q,t)\propto \frac{e^{-\frac{V(q)}{T}}}{t^\alpha}$ and  $\hat{f}(q,t) \approx \frac{e^{-\frac{q^2}{4D_{eff}t}}}{\sqrt{4\pi D_{eff}t}}$ for the diffusive length scales, $q << 2D_{eff}t$
and $q\approx 2D_{eff}t$, respectively with  $\alpha > 0$. For further details, one may go through Ref. \ cite {eli}. Here
using $\hat{f}(q,t)$, information entropy and other related properties were calculated.  However, keeping in mind the note after Eq.(\ref{eqm5b}), we deal with the equilibrium solution of (\ref{eqm5b}) to fulfill the major objective of the present study.

Furthermore, a universal relation between kinetic and entropy was established in Ref. \ cite {sor} with the distribution like

\begin{equation}
\hat{f}(q,t) = \frac{e^{-\frac{q^2}{4D^\prime t}}}{\sqrt{4\pi D^\prime t}} 
\label{eqm5e}
\end{equation}

\noindent
where the particle starts the journey from $q=0$ and $D^\prime$ is the effective relevant diffusion coefficient. It is applicable if $t$ is greater than the relaxation time. Here, it is noted that due to the central limit theorem, the above distribution function is applicable in most materials. In other words,
this is applicable for a sufficiently coarse-grained scale in space, the motion of the particle $q(t)$ takes the form of single ``hopping'' events which are independent of each other and equally distributed. A meaningful stationary state is not possible with  $\hat{f}(q,t)$ and  thus, the universal relation is between the kinetics and the information entropy.
 
We are now back to the present context. The solution of the Fokker-Planck equation (\ref{eqm5b}) at long time can be read as

\begin{equation}
f(q)=\frac{e^{-\frac{V(q)}{T}}}{{\langle e^{-V(q)/T}\rangle}_L} 
\label{eqm6}
\end{equation}
\noindent
Now, one can easily check that the above distribution function gives zero current, as we expect from the second law of thermodynamics, and thus, the check helps to convince us that it is a Boltzmann distribution function.
Then, the thermodynamic entropy($S$) of the system can be read as

\begin{equation}
S=ln{Q}+\frac{E}{T}  \; \; \;.
\label{eqm8}
\end{equation}

\noindent
where $Q=\int_{-L/2}^{L/2} e^{-V(q)/T}dq$ and $E=\langle V(q)\rangle$. Here, $Q$ can be identified as the partition function of the system, and
$E$ is the average energy of the system. 

\subsection{Microscopic derivation of the Adam-Gibbs relation}

     It is well known that the Adam-Gibbs relation seems to be valid experimentally when the temperature, as well as the entropy of the system, is very low. At this condition, one may assume that the particles are moving around the bottoms of wells of the potential energy field. Then, effectively, it can be expressed as
\begin{equation}
V=V(q_{min})+\frac{k}{2}(q-q_{min})^2 \; \; \;,
 \label{eqm9}
\end{equation}
where $V(q_{min})$ is the potential energy at the bottom of the well and $k=\frac{V_0}{2}(\frac{2\pi}{L})^2$. Using this relation into Eq.(\ref{eqm6}), we obtain the renormalized distribution function,

\begin{equation}
f(q)=\frac{1}{\sqrt{\pi\epsilon}} e^{-(q-q_{min})^2/\epsilon} 
\label{eqm10}
\end{equation}
\noindent
where $\epsilon=\frac{2T}{k}$. The thermodynamic entropy corresponding to Eq.(\ref{eqm10}) is given by

\begin{equation}
S=-\int dq f(q) ln f(q)=ln{\sqrt{\pi\epsilon}}+\frac{1}{2}  \; \; \;.
\label{eqm11}
\end{equation}
\noindent
One may obtain the above relation directly from Eq.(\ref{eqm8}) using Eqs.(\ref{eqm9}-\ref{eqm10}). However,  $\epsilon$ may be related to the effective diffusion coefficient. Thus, the above equation is an implicit relation between the diffusion coefficient and the entropy.

Since the entropy can not be negative, the above relation implies a critical value of $\epsilon$ for which it is zero. The critical value  $\epsilon_c$ can be read as 

\begin{equation}
\epsilon_c=e^{-(1+ln\pi)/2}=0.11  \; \; \;.
\label{eqm12}
\end{equation}
\noindent
The critical temperature $T_c$ corresponding to this is given by $T_c=\frac{k\epsilon_c}{2}$, which is proportional to the curvature of the potential energy function at the optimum points. In other words, it characterizes the boundness of the system. 
This temperature can be interpreted as a melting temperature.  However, with these definitions of $\epsilon_c$ and $T_c$, Eqs.(\ref{eqm11}) can be rearranged as 

\begin{figure}[!htb]
\includegraphics[width=9cm,angle=0,clip]{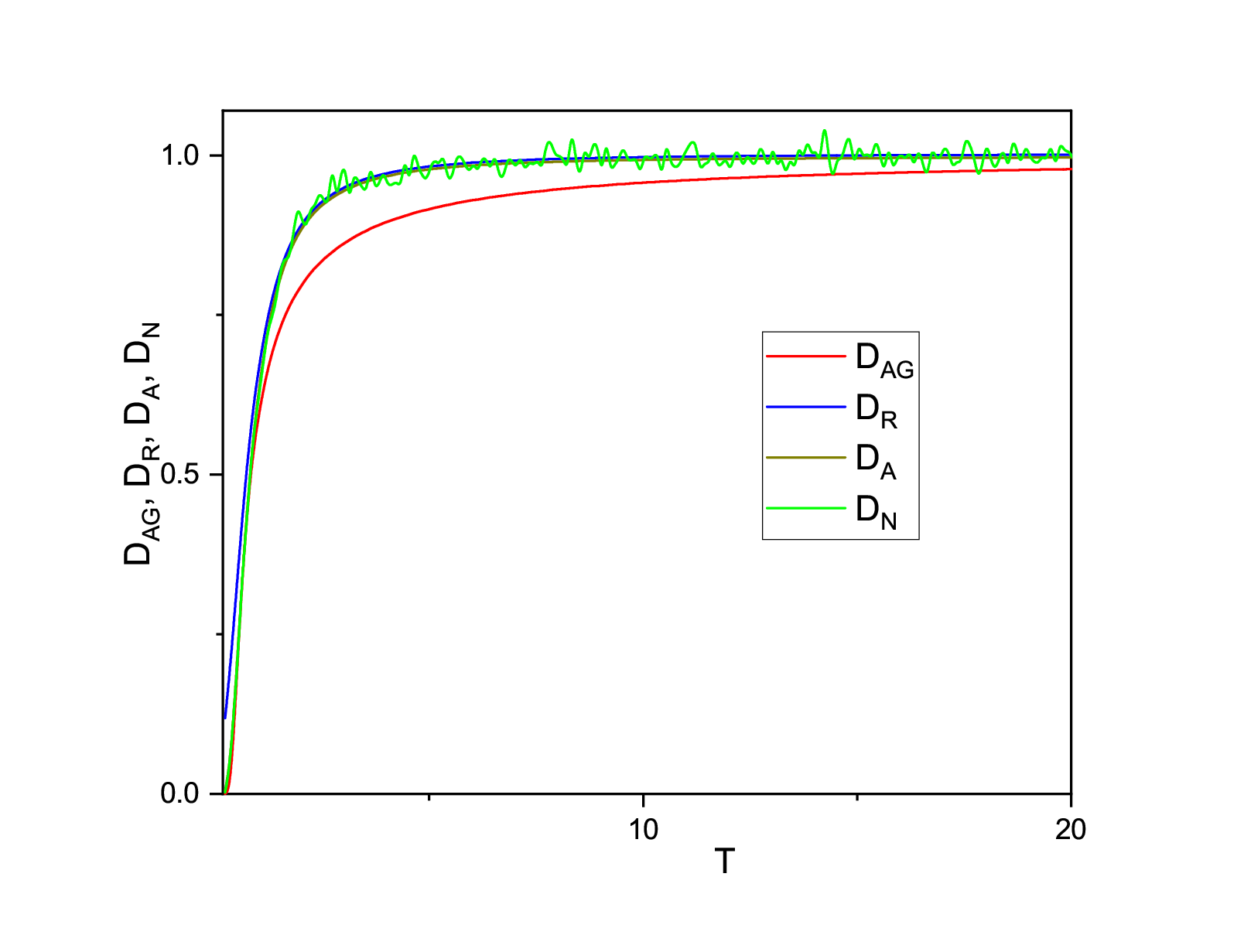}
\caption{ Demonstration of variation of different kinds of dimensionless diffusion constants($D_{AG}, D_N, D_A$, and $D_R$) with temperature
for the parameter set, $V_0=1.0, L=\pi$ and $\eta=5.0$. $D_{AG}, D_N, D_A$ and $D_R$ are defined in Eq.(\ref{eqm14}), Eq.(\ref{eqm16}), Eq.(\ref{eqm23a}) and (\ref{eqm24}), respectively.
(Units are arbitrary)}.;.
\label{Fig.1}
\end{figure}

\begin{equation}
2S=ln{\frac{T}{T_c}}   \; \; \;.
\label{eqm12a}
\end{equation}
\noindent
Then, multiplying both sides of the above equation by $-T$ we have

\begin{equation}
-\frac{ST}{b}=-Tln{\frac{T}{T_c}}   \; \; \;.
\label{eqm12b}
\end{equation}
\noindent
where $b=0.5$. Keeping in mind to reach the Adam-Gibbs relation, we rearrange the above equation as

\begin{equation}
e^{-\frac{b}{TS}}=e^{-\frac{1}{Tln{\frac{T}{T_c}}}}  \; \; \
\label{eqm13}
\end{equation}
\noindent
The connection between the entropy and the diffusion coefficient in  the Adam-Gibbs relation implies that the dimensionless
Adam-Gibbs diffusion coefficient ($D_{AG}$) is a only function of temperature. Then, one may identify the right hand side of the above equation
as 

\begin{equation}
D_{AG}=e^{-\frac{1}{Tln{\frac{T}{T_c}}}} \; \; \;.
\label{eqm15}
\end{equation}
\noindent
It characterizes the nature of the system in terms of $T_c$. Then one may write Eq.({\ref{eqm13}}) in the form of the Adam-Gibbs relation, 

\begin{equation}
D_{AG}=a e^{-\frac{b}{TS}} \; \; \;.
\label{eqm14}
\end{equation}
\noindent
where $a=1$. We now must test the validity of the relation (\ref{eqm14}). 
For this purpose, we use the Heun’s method \cite{toral} to solve Eq.(\ref{eqm1}) numerically with an integration step length of $0.005$. It is a stochastic version of the Euler method, which reduces to the second order Runge–Kutta method in the absence of the noise. A detail of this method was given in Ref. \ cite {shrabai}. The reliability of the method is well justified in \cite{bag6}. Solving Eq.(\ref{eqm1}) we calculate the dimension less diffusion coefficient($D_N$) based on the following standard relation,
\begin{equation}
D_N=\frac{\lim_{t\rightarrow \infty}\frac{\langle (q(t)-\langle q \rangle(t))^2\rangle_t}{2t}}{D} \; \; \;.
\label{eqm16}
\end{equation}
Here, the averaging was done considering 100000 trajectories.
$D_{AG}$ in Eq.(\ref{eqm14})
is calculated by determining the entropy using the distribution function (\ref{eqm6}). The results are compared in Fig.1. It shows that there is a very good agreement between the theory and the numerical experiment at low temperatures when the motion seems to be bounded in the well. The relation (\ref{eqm14}) is one of the key results of the present study. To the best of my knowledge, there is no report regarding the microscopic derivation of it. 

Now, one may generalize the relation  (\ref{eqm14}) for a $d$-dimensional space. Let  $V({\bf q}=\sum_{i=1}^{d}\hat{q}_iq_i)$ is an effective periodic potential in this space with a translational vector, ${\bf L}=\sum_{i=1}^{d}\hat{q_i}\frac{L}{\sqrt{d}}$. $q_i$ is the projection of the position vector ${\bf q}$
along $i-$th axis with unit vector $\hat{q}_i$. The unit vectors are orthogonal to each other. Thus, $L$ is the period in space as well as the length of the translational vector. $V({\bf q})$ may have many forms\cite{kl,kl1,kl2}. For example,

\begin{equation}
V(q_1,q_2)=V_0[\cos (2\pi q_1/L) + \cos (2\pi q_2/L)]   \; \; \;.   
\label{eqm3a}
\end{equation}
\noindent
This isotropic potential energy field in 2-dimensional space was used in Refs. \ cite {kl,kl1}. 

For a $d$-dimensional space, Eq.(\ref{eqm9}) can be written as

\begin{equation}
V=V(\{q_{i,min}\})+\frac{k}{2}\sum_{i=1}^{d}(q-q_{i,min})^2 \; \; \;,
\label{eqm9a}
\end{equation}
where $V(\{q_{i,min}\})$ is the potential energy at a point $\{q_{i,min}\}$ which corresponds to the bottom of a $d$-dimension well and we assume that the potential energy field is an isotropic in $d$-dimensional space. Then one may write that
$k_i=\frac{\partial^2 V}{\partial q_i^2}=k$ at the same point. Then, the Adam-Gibbs relation for a $d$-dimensional space is given by
 
\begin{equation}
D_{AG}=a e^{-\frac{b}{TS_d}} \; \; \;.
\label{eqm14a}
\end{equation}
where  $a=1$ and $b=\frac{d}{2}$. $S_d$ is the entropy in this space, and it is related to S in one dimension by $S_d=Sd$. 

To test the validity of this general relation(\ref{eqm14a}), we consider the potential energy field as given by Eq.(\ref{eqm3a}). For this Eq. (\ref{eqm6}) becomes

\begin{equation}
f(q_1,q_2)=\frac{e^{-\frac{V_1(q_1)}{T}}e^{-\frac{V_2(q_2)}{T}}}{{\langle e^{-V_1(q_1)/T}\rangle}_L{\langle e^{-V_2(q_2)/T}\rangle}_L} 
\label{eqm6a}
\end{equation}
\noindent
where $V_1(q_1)=V_0\cos (2\pi q_1/L), V_2(q_2)=V_0\cos (2\pi q_2/L),  {\langle e^{-V_1(q_1)/T}\rangle}_L=\frac{1}{L}\int_{-L/2}^{L/2} e^{-V_1(q_1)/T}dq_1$ and  ${\langle e^{-V_2(q_2)/T}\rangle}_L=\frac{1}{L}\int_{-L/2}^{L/2} e^{-V_2(q_2)/T}dq_2$.  
Then, the thermodynamic entropy($S_2$) of the system can be read as

\begin{equation}
S_2=2S  \; \; \;.
\label{eqm8a}
\end{equation}

\noindent
where $S=ln{Q_1}+\frac{E_1}{T}=ln{Q_2}+\frac{E_2}{T}, Q_1=\int_{-L/2}^{L/2} e^{-V_1(q_1)/T}dq_1=Q_2=\int_{-L/2}^{L/2} e^{-V_2(q_2)/T}dq_2$,
and $E_1=\langle V_1(q_1)\rangle=E_2=\langle V_2(q_2)\rangle$. Thus for the potential energy field (\ref{eqm3a}), Eq.(\ref{eqm14a}) becomes

\begin{equation}
D_{AG}=a e^{-\frac{b}{TS_2}} \; \; \;.
\label{eqm14b}
\end{equation}
where $a= $ and $b=1$.

\begin{figure}[!htb]
\includegraphics[width=9cm,angle=0,clip]{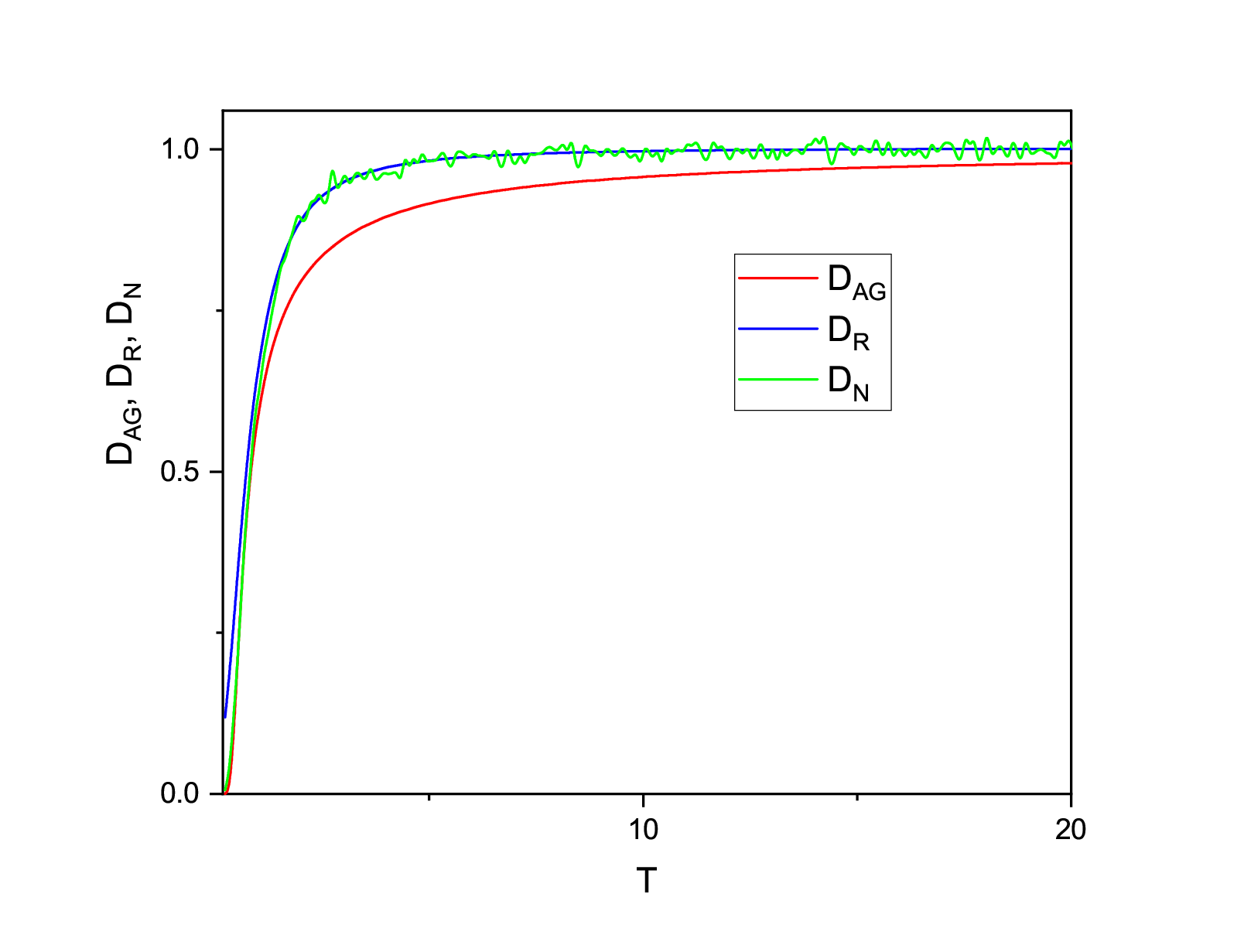}
\caption{ Demonstration of variation of different kinds of dimensionless diffusion constants($D_{AG}, D_N, D_R$) with temperature
for the parameter set, $V_0=1.0, L=\pi$ and $\eta=5.0$. $D_{AG}, D_N$ and $D_R$ are defined in Eq.(\ref{eqm14b}), Eq.(\ref{eqm16a}) and Eq.(\ref{eqm24b}), respectively.
(Units are arbitrary)}.
\label{Fig.2}
\end{figure}

Then, to calculate the effective diffusion coefficient numerically, we solve the following equations of motion as before,

\begin{equation}
\frac{dq_1}{dt}= \frac{2\pi V_0}{L\eta}\sin (2\pi q_1/L) +\xi_1(t)/\eta 
\label{eqm1a}
\end{equation}

\noindent
and

\begin{equation}
\frac{dq_2}{dt}= \frac{2\pi V_0}{L\eta}\sin (2\pi q_2/L) +\xi_2(t)/\eta 
\label{eqm1b}
\end{equation}

\noindent
$\xi_1(t)$ and  $\xi_2(t)$ are white Gaussian noise and they related to the damping strength as

\begin{equation}
\langle\xi_1(t)\xi_1(t^\prime)\rangle=\langle\xi_1(t)\xi_1(t^\prime)\rangle = 2k_BT\eta \delta(t-t^\prime) \; \; \;.
\label{eqm4a}
\end{equation}
\noindent
Solving the above equations of motion numerically we calculate the dimension less diffusion coefficient($D_N$) by the following standard relation,

\begin{equation}
D_N=\frac{\lim_{t\rightarrow \infty}\frac{\langle ({\bf r}(t)-\langle {\bf r} \rangle(t))^2\rangle_t}{4t}}{D} \; \; \;.
\label{eqm16a}
\end{equation}
\noindent
where ${\bf r}(t)=\hat{q}_1q_1+\hat{q}_2q_2$. Then, calculating $D_{AG}$ in Eq.(\ref{eqm14b}) with the relation (\ref{eqm8a}), we compare it with $D_N$ (as defined in the above equation) in Fig.2. Again it shows that there is an agreement between the theroy and the numerical experiment for low temperature when the motion seems to be bounded with in the well.

\subsection{Microscopic derivation of the Rosenfield relation}
 
We know that the Rosenfield relation is valid at high temperatures. Then, we need modification of the relation (\ref{eqm11}). One may do that
in the following way. Eq. (\ref{eqm11}) can be rearranged as

\begin{equation}
S-S_{id}=ln\sqrt{\frac{\pi\epsilon}{2L^2}}+ln\sqrt{\frac{\pi}{2L^2}}+ \frac{1}{2}-ln\sqrt{\frac{1}{2}}  \; \; \;.
\label{eqm17}
\end{equation}
\noindent
with the identification, $S_{id}=ln L$ as the ideal gas entropy. At relatively high temperature compared to the Adam-Gibbs limit, one assumes
from the above relation that

\begin{equation}
S-S_{id} \approx ln{\sqrt{\frac{\pi\epsilon}{2L^2}}}  \; \; \;.
\label{eqm18}
\end{equation}

\noindent
The above equation can be rearranged as

\begin{equation}
\frac{\epsilon}{2L^2}=ae^{b(S-S_{id})}  \; \; \;.
\label{eqm19}
\end{equation}
\noindent
with $a=1$ and $b=2$. 
Now, we have to determine $\epsilon$ properly. The quantity, $\frac{\epsilon}{2L^2}$ seems to be a qualitative measure of the ratio of size of nonideal system to the ideal one. In other words, $\frac{\epsilon}{2L^2}$ may be connected with $D_{eff}$ and $D$.

Then, we consider the dynamics. If   
$D_{eff}$ is the effective diffusion constant as defined in Eq.(\ref{eqm5d}),
\noindent
then one may assume that the mean square fluctuations of position, $\frac{\epsilon}{2} \approx \langle q^2\rangle =2D_{eff}t$ as suggested by the dynamics (\ref{eqm5c})(for details one may go through Ref.\cite{eli}). In this circumstance, the size ($L$) of the corresponding ideal gas system with uniform density may be as $L \approx \sqrt{2Dt}$.
Then one may write that
\begin{equation}
\frac{\epsilon}{2L^2} \approx D_A  \; \; \;.
\label{eqm23}
\end{equation}
\noindent
where
\begin{equation}
D_A=\frac{D_{eff}}{D}  \; \; \;.
\label{eqm23a}
\end{equation}

Now, identifying the dimensionless diffusion constant, $D_R=\frac{D_{eff}}{D}$ as a Rosenfield one, we have

\begin{equation}
D_R=ae^{b(S-S_{id})}  \; \; \;.
\label{eqm24}
\end{equation}
\noindent
We now calculate the right-hand side of this equation using the distribution function (\ref{eqm6}) and then compare it with the numerically  calculated the dimensionless diffusion coefficient(\ref{eqm16}). The results are demonstrated in Fig.1. It shows an excellent agreement between theory and numerical experiment for intermediate
to high temperature when the motion seems to be effectively unbounded. Thus, the above relation can be considered as an exact microscopic derivation of the Rosenfield relation.
Then, we notice that $b$ for the Rosenfield relation is inverse of the same in the Adam-Gibbs relation.

Now, following the previous subsection, one may generalize the above relation for a $d$-dimensional space, and then it can read as

\begin{equation}
D_R=ae^{b(S_d-S_{id,d})}  \; \; \;.
\label{eqm24a}
\end{equation}
\noindent
where $b=\frac{2}{d}$, $S_d=Sd$ and $S_{id,d}=dS_{id}$.  It is to be noted that a similar relation was derived in \cite{bagchi2} for an unbounded nonlinear system where ruggedness is distributed randomly. Here, it has been shown that $b=2$ for $d=1$ and $b=1$ for $d\geq 2$. The numerical experiment for this case
may confirm this prediction.

For the potential energy (\ref{eqm3a}), the above relation becomes

\begin{equation}
D_R=ae^{b(S_2-S_{id,2})}  \; \; \;.
\label{eqm24b}
\end{equation}
\noindent
with $a=1, b=1$ and $S_{id,2}=2 ln L$. $S_2$ is defined in Eq.(\ref{eqm8a}). Now calculating $D_R$, we compare it with $D_N$ (which is defined in Eq.(\ref{eqm16a})) in Fig.2. It shows very good agreement between the analytical result and the numerical one for intermediate to high temperature.

\subsection{Correspondence between the Adam-Gibbs and the Rosenfield relations}

It is to be noted here that Fig.1 demonstrates all the expected results, such as (i) at low temperature, the Adam-Gibbs relation is valid, (ii)
the Rosenfield relation is good for intermediate to high temperature, (iii) When the latter starts to work, then the former fails to be applicable. Thus, there is a transition temperature ($T_R$) at which two diffusion coefficient correspond to each other, and (iv) $D_A$ is exact for the entire range of temperature. Thus using Eqs.(\ref{eqm15}, \ref{eqm24}) $T_R$ can be determined by the relation

\begin{equation}
2T_Rln{\frac{T_R}{T_c}}(S(T_R)-S_{id})=-1 \; \; \;.
\label{eqm25}
\end{equation}
\noindent
$S(T_R)$ can be determined from (\ref{eqm11}). Thus, the above equation can be rearranged as

\begin{equation}
T_R ln{\frac{T_R}{T_c}} ln{\frac{T_R}{T_cL^2}}=-1 \; \; \;.
\label{eqm26}
\end{equation}
\noindent
It seems that finding of $T_R$ analytically from this nonlinear algebraic relation may not be easy. However, Eq.(\ref{eqm26}) suggests the following bound for $T_R$,

\begin{equation}
T_c < T_R < L^2T_c  \; \; \;.
\label{eqm27}
\end{equation}
From this relation, one may get a rough idea about the value of $T_R$. For example, the parameter set in Fig.1 suggests $T_R$ may lie  between
$0.22$ and $2.16$. For this figure, $T_R\approx 1$.

\section{Rosenfield-like relations for different contexts}

{\bf A thermodynamic system with free Brownian particles:}

    In the previous section, the main emphasis is to study $D_R$ for the introduction of a force field into the reference system with an ideal gas i. e., the assembly of non-interacting Broninam particles. Here, temperature and viscosity are the same for both nonideal and reference systems. One may
obtain different states of the ideal systems due to different temperatures or viscosities. Then, there may be a connection between the excess entropy and $D_R$. Another objective of this subsection is to address a relevant nontrivial issue (if any) in this context. For example,

{\bf Is it possible to recognize the configurational information entropy as a thermodynamic one for free Brownian particles after attainment of Maxwell's speed distribution?}

 It is apparent in both theoretical and numerical calculations in the previous section that at the high temperature limit, when the particle effectively does not experience an energy barrier to move,  the Rosenfield diffusion coefficient equals to one. At this limit, the distribution function can be read as

\begin{equation}
\hat{f}(q,t) = \frac{1}{\sqrt{4\pi Dt}}e^{-\frac{q^2}{4Dt}} \; \;,  
\label{eqm28}
\end{equation}
\noindent
which is corresponding to the following Langevin equation of motion

\begin{equation}
\frac{dq}{dt}=  \xi(t)/\eta 
\label{eqm29}
\end{equation}
\noindent
Then, the Fokker-Planck equation (\ref{eqm5}) becomes

\begin{equation}
\frac{\partial \hat{f}(q,t)}{\partial t}=D \frac{\partial^2 \hat{f}(q,t)}{\partial q^2} \; \;.
\label{eqm30}
\end{equation}
\noindent
This is the well-known Einstein's diffusion equation (with the solution (\ref{eqm28})), which is valid when the Brownian particles attend the Boltzmann distribution, and thus, the system is characterized by the thermodynamic temperature. It is to be noted here that using the solution of the Langevin equation (\ref{eqm29}), one can arrive at the distribution function (\ref{eqm28}) at long time when the average position becomes the initial position of the particle and the mean velocity is zero. At this limit, the solution of  equation (\ref{eqm29}) gives the celebrated Einstein's result, $\langle q^2\rangle_t= 2Dt$
which he obtained from the distribution function (\ref{eqm28}) with constant diffusion coefficient. It implies a peculiar kind of stationary state of the thermodynamic system whose volume as well as configurational Shannon entropy ($S_{Sh}$) increases with time at constant temperature and energy. The stationarity is due to the attainment of the  Maxwell-Boltzmann velocity distribution.  
This may be instructive in the inquiry of whether $S_{Sh}$ corresponds to the thermodynamic entropy. Then, one may look for a Hamiltonian, which at least may give an accurate description of the Brownian motion at equilibrium to calculate configurational thermodynamic entropy. The study in Ref. \ cite {liao} seems to be an approach in this direction. Here, the Hamiltonian $(H_R)$ (which mimics a harmonic potential with spring constant $m\eta^2$)  was written with respect to a reference Brownian trajectory. The spring constant suggests that it is a fictitious Hamiltonian, which is valid at equilibrium. If the number (N) of Brownian particles ($N\rightarrow \infty$) corresponds to a thermodynamic system, then the configurational entropy $S_{CTL}$ per particle moving in one dimension can be read for the Boltzmann-like distribution(it depends on the viscosity of the medium) with $H_R$ as  

\begin{equation}
S_{CTL}=ln{\sqrt{\frac{2\pi D}{\eta}}}+\frac{1}{2}  \; \; \;,
\label{eqm31}
\end{equation}

\noindent
where we have used $m=1$ and $D=\frac{T}{\eta}$. 
The thermodynamic entropy $S_{TV}$ in velocity space for the Hamiltonian is given by

\begin{equation}
S_{TV}=ln{\sqrt{2\pi D\eta}}+\frac{1}{2}=S_{CTL}+ln{\eta}  \; \; \;,
\label{eqm32}
\end{equation}
\noindent
It is to be noted here that $S_{TV}$ is independent of the viscosity of the medium, as we expect for thermodynamic entropy as well as Boltzmann distribution but $S_{CTL}$ depends on the same, and thus, one may say that it is a thermodynamic entropy-like quantity. Eq.(\ref{eqm32}) suggests that
the scaled configurational thermodynamic entropy, $S_{SCT}=S_{CTL}+ln{\eta}$ seems to be a true thermodynamic one.
Thus, the total TELQ ($_{TP}$) par particle in phase space is \cite{liao}

\begin{equation}
S_{TP}=ln{\frac{e D}{\hbar}}  \; \; \;,
\label{eqm33}
\end{equation}
\noindent
with $\hbar=\frac{1}{2\pi}$. This is equivalent to the situation when a Brownian particle is confined in a harmonic potential, $V(q)=\frac{\eta^2}{2}$, at equilibrium with the Boltzmann distribution

\begin{equation}
\hat{f}(p,q) =Z e^{-\frac{p^2}{2T}}e^{-\frac{\eta^2q^2}{2T}} \; \;,  
\label{eqm34}
\end{equation}
\noindent
where $p$ is the momentum of the particle, and $Z$ is the partition function. The reduced distribution function ($\hat{f}(q)$) in configuration space is given by 

\begin{equation}
\hat{f}(q) =\sqrt{\frac{\eta^2}{2\pi T}} e^{-\frac{\eta^2q^2}{2T}} \; \;,  
\label{eqm35}
\end{equation}
\noindent
The probability current for this is zero, which is not true for the actual system (\ref{eqm28}). Then the distribution function (\ref{eqm28}) with nonzero probability current 
suggests that thermodynamic entropy like quantity $S_{TCL}$ can be considered as a part of $S_{Sh}$ as follows

\begin{equation}
S_{Sh}(t)=-\int \hat{f}(q,t) ln \hat{f}(q,t) dq= ln{\sqrt{2\eta t}}+S_{CTL} \; \; \;.
\label{eqm36}
\end{equation}
\noindent
The contribution $S_{CTL}$ is for $t\ge\frac{1}{\eta}$. Thus, at the time ($t\approx\frac{1}{\eta}$ of attending Maxwell's speed distribution,
the configurational TELQ ($S_{CTL}$) is almost the same as that of the configurational Shannon information entropy. After this time, the
increase of $S_{Sh}$ as a signature of a free Brownian particle may be recognized as of fully irreversible origin as, like as isothermal free expansion of an ideal gas with zero net heat flux. In other words, the time and viscosity dependence of $S_{Sh}$  opposes considering $S_{Sh}$ as a thermodynamic entropy, although the system is characterized by the thermodynamic temperature with average energy, $\frac{T}{2}$.  It is to be noted here that the Hamiltonian formalisms\cite{charl,chandra, bender,chen} for dissipative system, including the Batman approach\cite{batman}, may not give the expected  Boltzmann distribution for both velocity and position of a Brownian particle.

We now show that one may obtain quantities like $S_{CTL}$ using the following shortcut ways.

(i) If we use the transformations,  $t=\frac{1}{\eta}$ and $x=\frac{q}{\sqrt{2}}$ into Eq.(\ref{eqm28}) then it becomes

\begin{equation}
\hat{f}_1(q) =\sqrt{\frac{k_1}{2\pi T}} e^{-\frac{k_1x^2}{2T}} \; \;.  
\label{eqm37}
\end{equation}
\noindent
It is the reduced Boltzmann distribution for the following Hamiltonian,
\begin{equation}
H_1(p,x) = \frac{p^2}{2}+\frac{k_1}{2}x^2    \; \;,  \label{eqm38}
\end{equation}
\noindent
with the fictitious spring constant, $k_1=\eta^2$. This is quite similar to $H_R$ in \cite{liao}, and thus we obtain configurational
thermodynamic entropy like quantity $S_{CTL1}$ for the distribution function (\ref{eqm37}) is equal to $S_{CTL}$. But the origin of the equilibrium states for the two cases is different. For $S_{CTL}$, we have already discussed above. The equilibrium state for the other case appears as a result of coupling of
$H_1(p,x)$ to a thermal bath. We consider this issue after writing possible fictitious Hamiltonians.

\noindent
(ii) We now use the transformation, $x=\frac{q}{\sqrt{2t}}$ or $x=\sqrt{\frac{\eta}{2t}}q$ into Eq.(\ref{eqm28}). Then, we obtain the following reduced distribution functions  and Hamiltonians 

\begin{equation}
\hat{f}_2(q) =\sqrt{\frac{k_2}{2\pi T}} e^{-\frac{k_2x^2}{2T}} \; \;., 
\label{eqm39}
\end{equation}

\begin{equation}
\hat{f}_3(q) =\sqrt{\frac{k_3}{2\pi T}} e^{-\frac{k_3x^2}{2T}} \; \;,  
\label{eqm40}
\end{equation}

\begin{equation}
H_2(p,q) = \frac{p^2}{2}+\frac{k_2}{2}x^2    \; \;,  
\label{eqm41}
\end{equation}

\noindent
and

\begin{equation}
H_3(p,q) = \frac{p^2}{2}+\frac{k_3}{2}x^2    \; \;,  
\label{eqm42}
\end{equation}

\noindent
where $k_2=\eta$ and $k_3=1$. The configurational TELQ for the distribution functions (\ref{eqm39}) and (\ref{eqm40}) is given

\begin{equation}
S_{CTL2}=ln{\sqrt{2\pi D}}+\frac{1}{2}  \; \; \;,
\label{eqm43}
\end{equation}

\noindent
and
\begin{equation}
S_{CTL3}=ln{\sqrt{2\pi D\eta}}+\frac{1}{2}  \; \; \;,
\label{eqm44}
\end{equation}
\noindent
These are related by
\begin{equation}
S_{CT2}=S_{CT3}-ln{\sqrt{\eta}}  \; \; \;,
\label{eqm45}
\end{equation}
\noindent
Similarly, $S_{CTL2}$ is related to $S_{CTL1}$ as well as $S_{CTL}$ by
\begin{equation}
S_{CTL}=S_{CTL1}=S_{CTL2}-ln{\sqrt{\eta}}  \; \; \;,
\label{eqm46}
\end{equation}
 
\noindent
(iii) Finally, the distribution function (\ref{eqm28}) itself can be identified as the reduced Boltzmann distribution function in configuration space as
\begin{equation}
\hat{f}_4(q) =\sqrt{\frac{k_4}{2\pi T}} e^{-\frac{k_4q^2}{2T}} \; \;.  
\label{eqm47}
\end{equation}
where $k_4=\sqrt{\frac{\eta}{2t}}$. Here, $t\ge \frac{1}{\eta}$. It corresponds to the state of the destination. Then, one may consider $t$  as a parameter for a system whose Brownian motion can be described by coupling the following Hamiltonian for the system,
\begin{equation}
H_4(p,q) = \frac{p^2}{2}+\frac{k_4}{2}x^2    \; \;,  
\label{eqm48}
\end{equation} 
\noindent
with the Hamiltonian corresponding to the thermal bath. Thus, $k_4$ is a constant in the above Hamiltonian. To avoid any confusion, one may represent the destination time $t$ by $\tau$ and $k_4=\frac{\eta}{2\tau}$. The viscosity dependence of the fictitious force constant is qualitatively expected;
as it becomes higher, the motion seems to be more confined in space for a given time. Similarly, the time dependence of $k_4$ is expected.
As the time increases, the distribution becomes flatter with decreasing entropy production $(\frac{dS_{Sh}}{dt}= \frac{1}{2t})$ and then expansion of the space seems to be easy. Now, using the Hamiltonian (\ref{eqm48}), one may be interested to have the answer to the following basic question.

{\bf Is the correspondence between the thermodynamic entropy and the Shannon information entropy a singular feature at equilibrium?}

The configuartional thermodynamic like entropy $S_{CTL4}$ associated with the distribution function (\ref{eqm47}) is given by
\begin{equation}
S_{CT4}= ln{\sqrt{\frac{2\pi T}{k_4}}}+\frac{1}{2}= S_{Sh}(t)  \; \; \;.
\label{eqm49}
\end{equation}
Now, one may feel good as we have at least a correspondence between the thermodynamic entropy-like quantity and the Shannon information entropy. $S_{CTL4}$
seems to be an exact one for the period, $\tau \ge t \ge \frac{1}{\eta}$. We elaborate it considering the dynamics.
The Langevin equations of motion corresponding to $H_4(p,q)$ is given by
 
\begin{equation}
\frac{dq}{dt}= p  
\label{eqm50}
\end{equation} 
  
\noindent
and 

\begin{equation}
\frac{dp}{dt}= -k_4 q -\eta p +\xi(t) \; \;. 
\label{eqm51}
\end{equation} 

\noindent
At overdamped limit Eq.(\ref{eqm51}) becomes

\begin{equation}
\frac{dq}{dt}= -\frac{k_4}{\eta} q +\xi(t)/\eta \; \;. 
\label{eqm52}
\end{equation} 
\noindent
According to these equations of motion, the respective trajectories seem to be different from the case with the actual Hamiltonian,  
$H(p) = \frac{p^2}{2}$. But it is not a serious point. At the early stage of dynamics, $q$ is small, and then its product with very small quantity, 
$k_4$ is a negiligible perturbation. Thus, the ensemble average trajectory and the mean square fluctuations of the observables according to the above equation of motion may correspond to the experimental  results. Then, we consider the aged dynamics ($\tau \ge t \ge \frac{1}{\eta}$). At this regime, the average position equals to initial position and the average velocity is zero for both the Hamiltonians. Mean square fluctuations of observables are also the same for both Hamiltonians. In other words, by mapping the original problem into the Hamiltonian description (\ref{eqm48}), we mean that the equilibrium state of the fictitious harmonic osscillator exactly correponds to the instantanious state at time $t=\tau$ of the Brownian particles and thus the correspondence in Eq.(\ref{eqm49}) leads to identifying $S_{Sh}$ as a thermodynamic entropy-like quantity.
Similarly, $S_{CTL}$ as well as $S_{CTL1}$ is almost equal to $S_{Sh}$ for $t=\frac{1}{\eta}$. $S_{CTL2}$ and $S_{CTL3}$ are related to $S_{CTL}$. However, by the nature of absorbing of time in the Hamiltonians,  $H_4(p,q)$ as well as $S_{CTL4}$ has a distinct feature to capture the experimental situation, as well as $S_{Sh}$ by the relation $t=\tau$. $H_4(p,q)$ suggests that the volume and the partition function-like quantity
diverge with time. It is to be noted here that $H_R$  also fails in this context.  
Furthermore, for $t=\infty$, $H_4(p,q)$ reduces to the actual Hamiltonian, $H(p)$, when the size of the system is the same as that of the container in which Brownian particles are uniformly distributed in space as like an ideal gas. Then, equilibrium appears in both space and velocity. At these conditions, 
$e^{-\frac{q^2}{4Dt}}\approx 1$ and Eq.(\ref{eqm28}) becomes

\begin{equation}
\hat{f}(q) = \frac{1}{L} \; \;,  
\label{eqm53}
\end{equation}
\noindent
where $L$ is the size of the one-dimensional container. Then, partition function, $Z=L$, and configurational thermodynamics entropy $S_{CT}$ is  
given by

\begin{equation}
S_{CT}= ln L \; \; \;.
\label{eqm54}
\end{equation}
Then, we check that the Shannon entropy ($S_{Sh}$) for this case is the same as that of thermodynamic entropy, i.e.,
\begin{equation}
S_{Sh}=-\int \frac{1}{L} ln\frac{1}{L} dq=S_{CT}= ln L   \; \; \;.
\label{eqm55}
\end{equation}
\noindent
As we expect. Thus, $H_4(p,q)$ gives an important result that the correspondence between the thermodynamic entropy and the Shannon information entropy is not a singular feature at equilibrium. Before equilibrium, $S_{Sh}$ is the same as that of a thermodynamic entropy-like quantity.

Thus, the present study provides a significant justification for the postulate that the Shannon entropy at the steady state may be considered like a thermodynamic entropy. This postulate plays a crucial role in the framework of stochastic thermodynamics\cite{sasa,seifert}.

We now mention one more issue. The average energy at long time for all the fictitious  Hamiltonians including $H_R$ is
\begin{equation}
E = \frac{T}{2}+ \frac{T}{2}  \; \;,  
\label{eqm56}
\end{equation}
\noindent
It has an extra term compared to the case $H(p) = \frac{p^2}{2}$ for which $E =\frac{T}{2}$. Then,  one may consider
the following Hamiltonian
\begin{equation}
H_i(p,q) = \frac{p^2}{2}+k_i\frac{q^2}{2} - \frac{T}{2}  \; \;, i=1,2,3,4 
\label{eqm57}
\end{equation}
\noindent
to bypass this inconsistency without affecting the desired results.

{\bf The Rosenfield-like relation for the Shannon information entropy} 

         In the previous section, we found that the Rosenfield relation connects a dimensionless diffusion coefficient with the excess thermodynamic entropy in configuration space. Then, one may look for a similar kind of relation for the Shannon information entropy with the distribution function (\ref{eqm28}). The diffusion constant which appears in Eq.(\ref{eqm28}) is related to the Shanon entropy(\ref{eqm36}) in configuration space by

\begin{equation}
Dt=a e^{b(S_{Sh}-\frac{1}{2})} \; \; \;.
\label{eqm58}
\end{equation}         
with $a=\frac{1}{4\pi}$ and $b=2$. In $d$-dimension space with isotropic diffusion constant, this relation can be written as         

\begin{equation}
Dt=a e^{\frac{2}{d}(S_{Shd}-\frac{d}{2})} \; \; \;.
\label{eqm59}
\end{equation} 
\noindent
where $S_{Shd}=dS_{Sh}$. If the diffusion constant is different for two cases of different thermal environment, such as $D=\frac{T}{\eta}$ and $D^o=\frac{T^o}{\eta^o}$, then, one may write the Rosenfiled-like relation(RLR) for the Shannon information entropy as

\begin{equation}
D_{RL}=a e^{b(S_{Shd}-S_{Shd}^o)}  \; \; \;.
\label{eqm60}
\end{equation}         
with $D_{RL}=\frac{D}{D^o}$, $a=1$, $b=\frac{2}{d}$ and $S_{Shd}^o$ is the information entropy for free Brownian motion in a thermal environment with temperature $T^o$ and viscosity coefficient $\eta^o$. Similarly, Eq.(\ref{eqm5e}) which  is applicable in most materials suggests Eq.(\ref{eqm60}) with $D_{RL}=\frac{D^\prime}{{D^\prime}^o}$. The relation like the above was derived in \cite{eli} using Eq. (\ref{eqm5c}).
Furthermore, if there is a motion with anomalous diffusion, then $t$ in Eqs. (\ref{eqm5e}) and (\ref{eqm28}) will be replaced by $t^\alpha$, where $\alpha$ is a constant. Then, the above relation is applicable. 

Till now, we consider a free particle or an unbounded nonlinear potential energy field. Then, we show that keeping fixed temperature and viscosity
of the thermal bath, $D$ can be different from $D^0$ in the presence of a magnetic field.

Following the system reservoir model\cite{shrabai}, the relevant Langevin equation of motion can be written as

\begin{equation}
\dot{x}=\dfrac{\Omega}{\eta}\dot{y}+\dfrac{1}{\eta}f_{x}
\label{od8}
\end{equation}

\noindent

\begin{equation}
\dot{y}=-\dfrac{\Omega}{\eta}\dot{x}+\dfrac{1}{\eta}f_{y}
\label{od9}
\end{equation}

\noindent
Here $\Omega=qB/m$ is the cyclotron frequency for a Brownian particle with mass $m$ and charge $q$. $B$ is the applied magnetic field. In Eqs.(\ref{od8}-\ref{od9})  $f_{y}$ and $f_{y}$ are independent white Gaussian noises, and they are related to the viscosity coefficient
by the relation,

\begin{equation}
\langle f_x(t) f_x(t^\prime)\rangle = \langle f_x(t) f_x(t^\prime)\rangle = 2k_BT\eta \delta(t-t^\prime)  
\label{oda}
\end{equation} 

Using Eq.(\ref{od9}) into Eq.(\ref{od8}) we 

\begin{equation}
\dot{x}=-\dfrac{\Omega^{2}}{\eta^{2}}\dot{x}+\dfrac{\Omega}{\eta^{2}}f_{y}+\dfrac{1}{\eta}f_{x}
\label{od10}
\end{equation}

\noindent
Similarly, we can write

\begin{equation}
\dot{y}=-\dfrac{\Omega^{2}}{\eta^{2}}\dot{y}+\dfrac{\Omega}{\eta^{2}}f_{y}+\dfrac{1}{\eta}f_{y}
\label{od11}
\end{equation}

Eq.(\ref{od10}) can be rearranged as

\begin{equation}
\dot{x}=\zeta(t)
\label{od13}
\end{equation}

\noindent
with
\begin{equation}
\zeta\left(t\right)=\dfrac{\Omega}{\eta^{2}+\Omega^{2}}f_{y}+\dfrac{\eta}{\eta^{2}+\Omega^{2}}f_{x} 
\label{od14}
\end{equation}
\noindent
Thus $\langle\zeta (t)\rangle=0$ and 

\begin{equation}
\langle\zeta(t)\zeta(t^\prime)\rangle=2D\delta(t-t^\prime)
\label{od15}
\end{equation}
 
\noindent 
with
\begin{equation}
D =  \dfrac{\eta T}{\eta^{2}+\Omega^{2}} \; \;.
\label{od16}
\end{equation}

Now, following the above calculation, we can write
\begin{equation}
S_{Shx}=\ln\sqrt{4\pi Dt}+\frac{1}{2}=S_{Shy} 
\label{od17}
\end{equation}
\noindent
Here, $S_{Shx}$ are $S_{Shy}$ are the Shannon information entropy for the equations of motion (\ref{od10}-\ref{od11}). Then we have the Rosenfield-like relation (\ref{eqm60}) with $D^0= \dfrac{T}{\eta}$,  $S_{Shd}=S_{Shx}+S_{Shy}$, $S_{Shx}=S_{Shy}$ and $d=2$.

{\bf The Rosenfield-like relation for the entropy analogous to the thermodynamic one:} 
         The relation between the diffusion constant and the configurational thermodynamic entropy-like quantity suggests the RLR
(\ref{eqm52}) with the same $b=\frac{2}{d}$ and the other constant $a$ depends on the viscosity coefficients which appear in  $D$ and $D^0$.
But the relation(\ref{eqm33}) between the total TDLE in phase space and the diffusion coefficient gives the following 
Rosenfield-like relation

\begin{equation}
D_{RL}=a e^{b(S_{TPd}-S_{TPd}^O)}  \; \; \;.
\label{eqm61}
\end{equation} 
\noindent
where $S_{TPd}=dS_{TP}$, $S_{TPd}^0=dS_{TP}^0$ $a=1$ and $b=\frac{1}{d}$. This relation between $b$ and $d$ was predicted in Ref. \ cite {liao}. It does not contradict the relation, $b=\frac{2}{d}$ for the configurational entropy for free Brownian motion or unbounded nonlinear systems. In other words, the relation, $b=\frac{1}{d}$, is a signature of the addition of the entropy in velocity space with the configurational entropy. Individually,
they follow the relation, $b=\frac{2}{d}$.         
          
\section{Conclusion}
          
          In the present paper, we consider the Brownian motion of particles in a periodic potential. Based on this, we calculate the configurational thermodynamic entropy of this system. This and other related calculations include the following major points.
          
\noindent
(i) There may be a critical temperature $T_c$ as a characteristic of a thermodynamic system, which is the lower bound to give nonzero
configurational thermodynamic entropy. It can be identified as a melting temperature.

\noindent
(ii) We derive the Adam-Gibbs relation, which seems to be very accurate around $T_c$. It is well-known from experiments that this relation is valid al low temperatures. Of course, low or high temperature should be defined with respect to $T_c$

\noindent
(iii) We also derive the Rosenfield field relation, which works well for intermediate to high temperatures. Thus, the validity range is consistent
with the empirical observations.

\noindent
(iv) Our calculation implies that there may be a correspondence between these relations, and then we have an equation to determine the transition temperature.

\noindent
(v) Considering a fictitious Hamiltonian, we show that the time-dependent configurational Shannon information entropy of a thermodynamic system (of  
Brownian particles) which is characterized by the absolute temperature, can not be recognised as a thermodynamic entropy. At best, one may identify
it as a thermodynamic entropy-like quantiy.               

\noindent
(vi) The description based on the fictitious Hamiltonian may lead to the  conclusion that the correspondence between the
Shannon information entropy and the thermodynamic entropy is not a singular feature at equilibrium. It is a continuation of the correspondence
between the information entropy and the thermodynamic entropy-like quantity. Thus, the present study seems to offer a compelling justification for the postulate that the Shannon entropy at steady state may be treated as a thermodynamic entropy. The postulate plays an important role in the field of stochastic thermodynamics.

\noindent
(vi) We generalize the  Rosenfield relation for both the information entropy and the thermodynamic entropy-like quantity.

\noindent
(vii) Finally, there is a constant $b$ (in the  Rosenfield and similar relations with configurational entropy) which is given
by $b=\frac{2}{d}$ for a $d$-dimensional unbounded nonlinear system or system of free particle.  The relation becomes, $b=\frac{1}{d}$
for the entropy in phase space\cite{liao}.

\noindent
\textbf{Acknowledgment}

\noindent
The $\;$ author is very pleased to recall the visit of Prof. $\;$ Biman Bagchi at Santiniketan on December, 2024. During this time, he was introduced to the field, the relation between the entropy and diffusion constant, with many important references by the professor. Then the discussions with  Prof. Bagchi motivated the author to search the fictitious Hamiltonians. At the same time, he and Prof. Kazuhiko Seki are heartily acknowledged for reading the paper and giving suggestions. Finally, the author is also heartily thankful to Mousumi Biswas, who helped him
by drawing the figures and editing the English in the paper.

\end{document}